\documentclass[runningheads]{llncs}
\usepackage[T1]{fontenc}

\usepackage{graphicx}
\usepackage[most]{tcolorbox}
\usepackage{amsmath,amsfonts}
\usepackage{xcolor}
\definecolor{added}{rgb}{0.8, 1, 0.8} 
\definecolor{removed}{rgb}{1, 0.8, 0.8} 

\lstdefinestyle{myStyle}{
    belowcaptionskip=1\baselineskip,
    breaklines=true,
    frame=single,
    basicstyle=\footnotesize\ttfamily,
    commentstyle=\itshape\color{purple!40!black},
    escapeinside={(*@}{@*)},
}
\usepackage{microtype}
\usepackage{balance}
\usepackage{graphicx}
\usepackage{textcomp}
\usepackage{xcolor}
\usepackage{color, colortbl}
\usepackage{caption}
\usepackage{subcaption}
\usepackage[utf8]{inputenc}
\usepackage{verbatimbox}
\usepackage{multirow}
\usepackage{makecell}
\usepackage{adjustbox}
\usepackage[scientific-notation=true]{siunitx}
\usepackage{todonotes}
\usepackage{balance}

\usepackage{ dsfont }
\usepackage{stmaryrd}

\usepackage[Algorithm,ruled]{algorithm}
\usepackage{url}
\usepackage[noend]{algpseudocode}
\usepackage{graphicx}

\usepackage{textcomp}
\usepackage{xcolor}
\usepackage{booktabs}
\usepackage{xspace}
\usepackage{enumitem}
\usepackage{listings}
\usepackage{pifont}
\usepackage{listings,multicol}
\usepackage{float}
\usepackage[utf8]{inputenc}
\usepackage{verbatim}
\usepackage{hyperref}
\usepackage{multirow}
\usepackage{makecell}
\usepackage{adjustbox}
\usepackage[scientific-notation=true]{siunitx}
\usepackage{balance}
\usepackage{amsmath,amsfonts}

\usepackage[Algorithm,ruled]{algorithm}
\usepackage{url}
\usepackage[noend]{algpseudocode}
\usepackage{graphicx}

\usepackage{textcomp}
\usepackage{xcolor}
\usepackage{booktabs}
\usepackage{xspace}
\usepackage{enumitem}
\usepackage{listings}
\usepackage{pifont}
\usepackage{fancyvrb}
\usepackage{framed}
\usepackage{xcolor}

\definecolor{codered}{HTML}{D9534F}
\definecolor{codegreen}{HTML}{5CB85C}
\definecolor{codeblue}{HTML}{0000FF}

\definecolor{m1}{RGB}{224, 250, 255}
\definecolor{m2}{RGB}{219, 224, 255}
\definecolor{m3}{RGB}{252, 255, 220}
\definecolor{m4}{RGB}{249, 214, 177} 

\usepackage{tikz}
\usetikzlibrary{shapes.geometric}
\usepackage{listings}
\usepackage{orcidlink}
\usepackage{pgfplots}
\usepackage[]{footmisc}

\tikzset{
    use bounding box relative coordinates/.style={
        shift={(current bounding box.south west)},
        x={(current bounding box.south east)},
        y={(current bounding box.north west)}
    },
}

\tikzstyle{normal} = [rectangle,text centered,draw=white]
\lstdefinestyle{myStyle}{
    captionpos=b,
    breaklines=true,
    frame=single,
    numbers=left,
    basicstyle=\scriptsize\ttfamily,
    commentstyle=\itshape\color{purple!40!black},
    identifierstyle=\color{blue},
    backgroundcolor=\color{gray!10!white},
    belowskip=0pt
}
\begin{document}
\title{Quantitative Symbolic Patch Impact Analysis}
%
%
\author{Laboni Sarker\orcidlink{0000-0002-4793-7859} \and Abdus Satter\orcidlink{0000-0002-2053-7150}  \and
Tevfik Bultan\orcidlink{0000-0003-2993-1215}   \\
\email{\{labonisarker, abdussatter, bultan\}@ucsb.edu}}
\authorrunning{L. Sarker  et al.}
\institute{University of California, Santa Barbara}

%
\maketitle              
\begin{abstract}

Traditional equivalence checking classifies programs as equivalent or non-equivalent, providing insufficient information for tasks like patch impact analysis where it is expected the patched version of the program to be non-equivalent to the original program. When two program versions are non-equivalent, determining \textit{under what conditions} they differ and \textit{what percentage of inputs} are affected remains an open challenge. In this work, we introduce quantitative partial equivalence analysis, an approach for assessing software patches by quantifying behavioral differences between the original (vulnerable) code and the patched code. Using symbolic analysis, we identify input conditions under which patched and original programs exhibit identical or divergent behaviors. Our approach refines non-equivalence by measuring the extent of behavioral divergence across the input domain. For efficient quantitative analysis of numerical domains, we propose a range-based search heuristic that provides a sound lower bound on equivalence. We demonstrate our approach on 90 CVE patches from widely used open-source projects (Linux, Qemu, FFmpeg), as well as on a Juliet Test Suite-based dataset containing programs with CWEs. Our results show that quantitative partial equivalence analysis effectively characterizes and quantifies patch impact. Additionally, experiments on the EqBench benchmark reveal five C program pairs that are mislabeled as equivalent, and we identify the input conditions under which their behaviors diverge.



\

\vspace*{-0.2in} 

\end{abstract}




\section{Introduction}
\label{sec:introduction}
\let\thefootnote\relax
\footnotetext{This material is based on research supported by NSF under grant 2008660, by ONR Contract No. N6833523C0019, Oceanit Laboratories, and by DARPA under Agreement No. HR0012590048. The U.S. Government is authorized to reproduce and distribute reprints for Governmental purposes notwithstanding any copyright notation thereon. The views and conclusions contained herein are those of the authors and should not be interpreted as necessarily representing the official policies or endorsements, either expressed or implied, of the U.S. Government.}

Two programs are considered non-equivalent even if the non-equivalence arises from only a single input within the entire input domain. 
Unlike being ``equivalent'', which implies equivalence across the entire input domain, being ``non-equivalent'' does not necessarily entail divergence over the entire domain~\cite{quantitaveSimilarityAnalysis,quantitativeNeqAnalysis}. For many applications, such as patch analysis, it is important to analyze the specific input conditions under which the two non-equivalent programs exhibit equivalent or differing behaviors. Additionally, quantifying the {\em level of equivalence} in such cases yields more refined insights into the relationship between the two programs.
In this paper, we focus on 
``partial equivalence'' where two programs behave equivalently for a subset of the input domain.



In patch analysis~\cite{patchSemanticAnalysis,patUntrack,SPAIN}, the expected case is non-equivalence, where the original code and the patched version are not expected to be equivalent, as the patch introduces modifications to remediate the vulnerability. However, patches rarely affect the entire input domain~\cite{ghanbari2022patch}. A naive remediation strategy may simply remove functionality by rejecting all inputs~\cite{spider} or concretize inputs to a fixed non-vulnerable value~\cite{black2018juliet,boland2012juliet}. In both cases, the original and patched programs are non-equivalent; however, such patches introduce non-equivalence over a significantly larger portion of the input domain. This observation highlights the need for a more refined and quantitative analysis of non-equivalence.



Although there has been prior work on equivalence analysis~\cite{ardiff,pasda,dse} and on quantification of software changes~\cite{quantificationSoftware1}, to the best of our knowledge, there has not been prior work on developing quantitative partial equivalence analysis techniques for assessing patch impact. In this paper, we formalize the notion of partial equivalence, propose techniques for deriving equivalence conditions for partially equivalent programs, quantify the fraction of the input domain over which programs behave equivalently or differently, and apply this analysis to assess patch impact. Our work builds on and extends prior work on differential symbolic execution~\cite{dse}, equivalence analysis~\cite{ardiff,pasda,kawaguchi2010conditional}, quantitative program analysis~\cite{quantificationSoftware1,probabilisticSE}, and patch analysis~\cite{spider,sawadogo2022sspcatcher,tian2012identifying}.

Our contributions in this paper include:
\vspace*{-0.08in}
\begin{itemize}
    \item Formalization of the concept of partial equivalence that refines the ``not-equivalent'' results in equivalence analysis, and quantitative extension of partial equivalence analysis with equivalence/non-equivalence conditions and corresponding percentages.
    \item A range-search based technique for numeric input domains for efficient quantitative partial equivalence analysis.
    \item Application of our quantitative partial equivalence analysis technique to patch impact analysis and experimental evaluation on a patch dataset that consists of patches collected from real-world applications~\cite{ffmpeg,qemu,linux} and Juliet Test Suite~\cite{boland2012juliet}.
    \item Application of our partial equivalence analysis technique on EqBench~\cite{eqBench}, a dataset for equivalence analysis, that identifies incorrectly labeled equivalent test cases in EqBench along with identifying the divergent input conditions.
\end{itemize}
The remainder of the paper is organized as follows. We discuss motivating examples in Section \ref{sec:overview}, formalization of partial equivalence and patch impact surface in Section~\ref{sec:partialEq}, quantitative techniques for partial equivalence in Section~\ref{sec:partialEqAnalysis}, discussion of patch dataset in Section~\ref{sec:patchBench}, implementation and experimental evaluation in Section~\ref{sec:experiment}, related works in Section~\ref{sec:relatedwork}, and conclusions in Section \ref{sec:conclusion}.


\vspace*{-0.08in}
\section{Motivating Examples}
\label{sec:overview}
\vspace*{-0.08in}





For vulnerabilities with available patches not yet deployed in dependent applications, attackers actively seek user-controlled inputs that can still trigger the underlying weaknesses~\cite{pulliainen2016linux,lefdal2022security}. Often, their goal is broader exploitation, such as privilege escalation or memory corruption, which may require chaining multiple vulnerabilities~\cite{kallenberg2014extreme,overviewLinuxVul,chen2011linux}. Attacks are more likely to succeed on vulnerabilities that are easily triggered, i.e., those affecting a large portion of the input domain. Beyond security concerns, patches or software updates that affect a wide input domain may require extra scrutiny from maintainers, as they can significantly alter program behavior (e.g., the Crowdstrike outage caused by an errant update~\cite{george2024trust}). Patches exhibiting greater non-equivalence also increase the risk of breaking existing assumptions and CI/CD pipelines~\cite{backwardCompatibility}, highlighting the need for more rigorous testing and validation.


The quantitative patch impact analysis techniques we present in this paper 1) identify the conditions under which a patched program diverges from its original version and 2) quantify the proportion of affected inputs. 



\begin{lstlisting}[caption=Linux: CVE-2012-2384, label={lst:LinuxOverflow189}, language=C, xleftmargin=5.0ex, firstnumber=1058, escapeinside={(*@}{@*)}]
struct drm_clip_rect *cliprects = NULL;
...(*@\setcounter{lstnumber}{1129}@*)
if (args->num_cliprects != 0) {
...(*@\setcounter{lstnumber}{1135}@*)
(*@\aftergroup\speciallstcolorAdd@*) +if (args->num_cliprects > (*@\aftergroup\endspeciallstcolor@*) (*@\textcolor{green!40!black}{UINT\_MAX}@*) (*@\aftergroup\speciallstcolorAdd@*)/ sizeof(*cliprects)) {(*@\aftergroup\endspeciallstcolor@*)
 (*@\aftergroup\speciallstcolorAdd@*)+ DRM_DEBUG("execbuf with %u cliprects",args->num_cliprects)(*@\aftergroup\endspeciallstcolor@*)
 (*@\aftergroup\speciallstcolorAdd@*)+ return -EINVAL; }(*@\aftergroup\endspeciallstcolor@*)
cliprects = kmalloc(args->num_cliprects * sizeof(*cliprects), GFP_KERNEL);
\end{lstlisting}
\vspace*{-0.1in}

Listing~\ref{lst:LinuxOverflow189} illustrates a patch where the userspace variable {\tt args->num\_cliprects} can cause out-of-bounds access. In the patch, an additional condition is added on this variable at line 1136, which is later used at line 1139. Before the update, the variable was only checked for being non-zero at line 1130. These changes make the original and patched code non-equivalent. Our analysis identifies the condition for non-equivalence as $(536870911 < \mbox{\tt args->num\_cliprects} \leq \mbox{\tt UINT\_MAX})$, where $\mbox{\tt UINT\_MAX} \div \mbox{\tt sizeof(*cliprects)} = 536870911$, quantifying the patch impact as 87.50\% of the input domain (i.e.,
87.50\% of inputs will behave differently for the original program and the patched program).
\begin{lstlisting}[caption=Linux: CVE-2010-4165, label={lst:numericError189_4165}, language=C, xleftmargin=5.0ex, firstnumber=61, escapeinside={(*@}{@*)}]
#define MAX_TCP_WINDOW		32767U
...(*@\setcounter{lstnumber}{2249}@*)
(*@\aftergroup\speciallstcolor@*)- if (val < 8 || val > MAX_TCP_WINDOW) {(*@\aftergroup\endspeciallstcolor@*)
(*@\aftergroup\speciallstcolorAdd@*)+ if (val < 64 || val > MAX_TCP_WINDOW) {(*@\aftergroup\endspeciallstcolor@*)
			err = -EINVAL;
			break; }
tp->rx_opt.user_mss = val;
\end{lstlisting}
\vspace*{-0.1in}

In Listing~\ref{lst:numericError189_4165}, line 2250 is removed and line 2251 is added. For the variable {\tt val}, our analysis detects non-equivalence when $8 \leq \mbox{\tt val} \leq 63$, affecting only 56 inputs out of $2^{32}$. Both Listings~\ref{lst:LinuxOverflow189} and~\ref{lst:numericError189_4165} are categorized as CWE-189 numeric errors~\cite{wang2020machine}. However, our patch impact analysis shows that the patch from Listing~\ref{lst:LinuxOverflow189} impacts a larger portion of the program behavior compared to Listing~\ref{lst:numericError189_4165}, along with identifying the non-equivalence conditions for each patch. 

\begin{lstlisting}[caption=FFmpeg: CVE-2013-0859, label={lst:ffmpeg_189_0859}, language=C, xleftmargin=5.0ex, firstnumber=249, escapeinside={(*@}{@*)}]
static int add_doubles_metadata(int count,
...(*@\setcounter{lstnumber}{254}@*)
(*@\aftergroup\speciallstcolor@*)- if (count >= INT_MAX / sizeof(int64_t))(*@\aftergroup\endspeciallstcolor@*)
(*@\aftergroup\speciallstcolorAdd@*)+ if (count >= INT_MAX / sizeof(int64_t) || count <= 0)(*@\aftergroup\endspeciallstcolor@*)
         return AVERROR_INVALIDDATA;
  if (bytestream2_get_bytes_left(&s->gb) < count * sizeof(int64_t))
\end{lstlisting}
\vspace*{-0.1in}

Another patch from FFmpeg is shown in Listing~\ref{lst:ffmpeg_189_0859} where a modification is made to fix out of array access by adding a constraint on the {\tt count} variable at line 255, which is later used in line 258. Our analysis shows that this update will impact 1 input for variable {\tt count} with the corresponding non-equivalence condition: $\mbox{\tt count} = 0$.

Based on our patch impact analysis for CWE-189 numeric errors, Listing~\ref{lst:LinuxOverflow189} warrants greater attention from software maintainers (for example, in terms of the amount of testing required) before deployment compared to Listings~\ref{lst:numericError189_4165} and~\ref{lst:ffmpeg_189_0859}, as it affects a significantly larger portion of the input domain.

\section{Partial Equivalence and Patch Impact Analysis} 
\label{sec:partialEq}
In this section, we formalize the concept of partial equivalence and discuss its application to patch impact analysis. We use a deterministic, terminating program model in which each input produces a single output. The term \emph{program} is used broadly to refer to a program, a function, or a code segment where input and output variables are identified.

\begin{definition}
A program $P$ is a total function from the domain of inputs to the domain of outputs, $P: \mathbb{D}_{I_P}  \rightarrow \mathbb{D}_{O_P} $, 
where 
$P(\overrightarrow{I}) = \overrightarrow{O}$ denotes that on input $\overrightarrow{I} \in \mathbb{D}_{I_P} $ the output of $P$ is $\overrightarrow{O} \in \mathbb{D}_{O_P} $.
Each input of $P$ is a vector of input parameters $\overrightarrow{I} = [i_1, i_2, \ldots , i_N]$ where $N$ denotes the number of input parameters.  
Similarly, the output of $P$ is a vector $\overrightarrow{O} = [o_1, o_2, \ldots , o_M]$ where $M$ is the total number of resultants of the program.
Input domain is $\mathbb{D}_{I_P}  =  D_{i_1}\times D_{i_2}\times  \ldots \times D_{i_N}$ where $D_{i_n}$ denotes the set of possible values of input $i_n~(1\leq n\leq N)$. Similarly, the output domain is $\mathbb{D}_{O_P}  =  D_{o_1}\times D_{o_2}\times \ldots \times D_{o_M}$ where $D_{o_m}$ defines  the set of possible values that output $o_m~(1\leq m\leq M)$ can take. 
\label{def:program}
\end{definition}



Given two programs $P_1, P_2$ with same input domains (where $\mathbb{D}_I=\mathbb{D}_{I_{P_1}} = \mathbb{D}_{I_{P_2}}$), let $\mathbb{D}_{\textit{eq}}  \subseteq \mathbb{D}_I$ (equivalence set) denote the set of inputs where the two programs generate the same output (i.e., $\overrightarrow{I} \in \mathbb{D}_{\textit{eq}}
\Leftrightarrow P_1(\overrightarrow{I})=
P_2(\overrightarrow{I}$)).
We define $\mathbb{D}_{\textit neq}$ (the non-equivalence set) as: 
$\mathbb{D}_{\textit neq} = \mathbb{D}_I - \mathbb{D}_{\textit eq}$.
Then, we assess the equivalence between two programs as follows: 
\begin{definition}
Given two programs $P_1, P_2$ 
with equivalence and non-equivalence sets, $\mathbb{D}_{\textit{eq}}$ and $\mathbb{D}_{\textit{neq}}$, respectively, we define: 
\begin{itemize}
\item \textbf{Equivalence:} 
$\mathbb{D}_{\textit{neq}}  = \emptyset$ 
\item \textbf{Total non-equivalence:}  $\mathbb{D}_{\textit{eq}}  = \emptyset$ 
\item \textbf{Partial equivalence:}
  $\mathbb{D}_{\textit{neq}}  \neq \emptyset \land \mathbb{D}_{\textit{eq}}  \neq \emptyset$ 
\end{itemize}
\label{def:equivalence}
\end{definition}

\begin{definition}
Given two programs $P_1$, $P_2$ with equivalence set  $\mathbb{D}_{\textit{eq}}$, the \textbf{equivalence condition} $F_{\textit eq}$, is a formula on input variables $N$, where,  
$\mathbb{D}_{\textit{eq}} = \llbracket F_{\textit{eq}} \rrbracket$.
\label{def:eqCondition}
\end{definition}

We apply these concepts to patch impact analysis as follows:

\begin{definition}
Given a vulnerable program $P_1$ and its patched version $P_2$, we define the \textbf{patch impact surface} 
as the non-equivalence set $\mathbb{D}_{\textit{neq}}$ of $P_1$ and $P_2$.
\label{def:patchimpactsurface}
\end{definition}
\noindent
I.e., patch impact surface is the set of all inputs where the patched version of the program behaves differently than the original version of the program. Furthermore, we can use the partial equivalence condition ($F_{\textit{eq}}$) to identify the input condition characterizing the patch impact surface, which is 
$\neg F_{\textit{eq}}$, since
$\mathbb{D}_{\textit{neq}} = \llbracket \neg F_{\textit{eq}} \rrbracket$.

Given a vulnerable program with a specific vulnerability, the attack surface of the program with respect to that vulnerability is defined as all inputs that trigger the vulnerability. For example, if vulnerability is captured by an assertion, the attack surface would be all inputs that cause the assertion violation.

Given a program $P_1$ and its patch $P_2$, the patch $P_2$ is considered secure with respect to the given vulnerability if its attack surface is empty. For any secure patch, its patch impact surface should subsume the attack surface of the original program. There always exists a trivial secure patch, for instance, one that halts immediately after receiving input, but such patches are useless because they destroy program functionality. Therefore, developers aim to select a patch whose impact surface matches the original attack surface, ensuring that only vulnerable inputs are modified while preserving program behavior for all other inputs.

In this paper, we present techniques to compute and quantify the patch impact surface, allowing us to measure the fraction of the input domain affected by a patch. A large patch impact surface could indicate that the patch modifies program behavior beyond what is necessary, potentially  breaking functionality. When a large impact surface is unavoidable due to a large attack surface (a large number of inputs trigger a vulnerability), extensive testing is justified because of the combined risk of functional changes and high exploitability. Hence, for all these reasons, ability to identify and quantify the patch impact surface is a crucial problem that we address with the quantitative symbolic partial equivalence analysis techniques we present.

\section{Quantitative Symbolic Partial Equivalence Analysis}
\label{sec:partialEqAnalysis}


In this section we first discuss symbolic partial equivalence analysis, followed by its extension to quantitative symbolic partial equivalence analysis.
Symbolic partial equivalence analysis starts with generating symbolic summaries~\cite{dse} of programs using extended symbolic execution.




\begin{definition}
Given a program  $P: \mathbb{D}_{I_P}  \rightarrow \mathbb{D}_{O_P} $, 
with  $N$ inputs and $M$ outputs, symbolic summary of program $P$, denoted as $S_P$, is a logical formula with $N+M$ free variables corresponding to the inputs and the outputs of the program $P$, where
\[
\overrightarrow{I}, \overrightarrow{O} \models \llbracket S_P \rrbracket \ \ 
\mbox{if and only if} \ \ 
P(\overrightarrow{I}) = \overrightarrow{O}
\]
\label{def:summary}
\vspace*{-10pt}
\end{definition}


Symbolic execution explores the execution paths in a program by representing input parameters symbolically and capturing constraints on those parameters for each path as a path constraint. For generating the symbolic summary of a program based on Definition~\ref{def:summary}, we also need to capture the outputs corresponding to each path and express constraints on the outputs as part of the path constraints of the program. Extended symbolic execution collects output constraints along with the input path constraints. The symbolic summary of a program generated by extended symbolic execution is a disjunction of path constraints (one for each path explored by symbolic execution) and is expressed as: 
\[ 
S \equiv C_1 \vee C_2 \vee ...\vee C_K
\]
where each path constraint $C_k (1\leq k\leq K)$ represents the constraints on input $\overrightarrow{I}$ in conjunction with constraints on the corresponding output $\overrightarrow{O}$ for path $k$. 

\begin{algorithm}[t]
\caption{{\sc EqChecker}($P_1, P_2$)\\
 Takes two programs $P_1$, $P_2$ and classifies the programs as equivalent $(T_{eq})$, totally non-equivalent $(T_{neq})$ or partially equivalent $(P_{eq})$.}
\label{proc:classifier}
\begin{footnotesize}
\begin{algorithmic}[1]
\State $S_1 \leftarrow $ \textsc{Summarize}($P_1$)
\State $S_2 \leftarrow $ \textsc{Summarize}($P_2$)
\If{ $\neg$ \textsc{IsSat}$(\neg (S_1 \Leftrightarrow S_2))$ 
 }
 \Comment if $S_1 \Leftrightarrow S_2$  is valid
      \State \Return  $T_{eq}$  
\EndIf
\If{$\neg$ \textsc{IsSat}$(S_1 \land S_2)$}
      \State \Return  $T_{neq}$  
\EndIf
\State \Return  $P_{eq}$ 
\end{algorithmic}
\end{footnotesize}
\end{algorithm} 
Algorithm~\ref{proc:classifier} classifies given input programs into three categories: equivalent ($T_{eq}$), totally non-equivalent ($T_{neq}$), or partially equivalent ($P_{eq}$), as defined in Definition~\ref{def:equivalence}. The $\textsc{Summarize}$ function within the algorithm generates symbolic summaries~\cite{dse}, while the $\textsc{IsSat}$ function checks the satisfiability of a given formula. 
Given programs $P_1$, $P_2$ and their symbolic summaries $S_1$, $S_2$ respectively, based on the Definitions~\ref{def:program},~\ref{def:equivalence} and~\ref{def:summary}, we have:
\begin{itemize}
\item $P_1$ and $P_2$ are equivalent if $S_1 \Leftrightarrow S_2$ is valid (line 3),
\item $P_1$ and $P_2$ are totally non-equivalent if $S_1 \land S_2$ is not satisfiable (line 5),
\item $P_1$ and $P_2$ are partially equivalent otherwise (line 7). 
\end{itemize}

Given the input domain $\mathbb{D}_I$ for a program, the size of the input domain is denoted as $|\mathbb{D}_I|$. The number of input values where the behavior of the two programs are equivalent (the size of the equivalent set) is denoted as $|\mathbb{D}_{\textit eq}|$. Then, the size of the non-equivalent set corresponds to:
\textbf{$|\mathbb{D}_{\textit neq}| = |\mathbb{D}_I| - |\mathbb{D}_{\textit eq}|$}.

\begin{definition}
Given an input domain $\mathit{\mathbb{D}_I}$ and the equivalent set $\mathit{\mathbb{D}_{\textit eq}}$, the {\em equivalence percentage} is the ratio of the size of equivalent set and the size of input domain which can be formulated as $\frac{\mathit{|\mathbb{D}_{\textit eq}|}}{\mathit{|\mathbb{D}_I|}} \times 100$. 
\label{def:percentage}
\end{definition}


If two given programs are partially equivalent ($P_{eq}$), our range-based search technique, which we present in the next section, provides (i) the equivalence condition and (ii) the equivalence percentage, i.e., the percentage of the input domain for which the given programs behave equivalently.

\subsection{Partial Equivalence Analysis with Range Search}
\label{subsec:rangeBasedSearch}



Partially equivalent programs are more likely to exhibit equivalent behavior for a continuous subdomain, i.e., a range, rather than for scattered values across the input domain. Based on this intuition, we propose heuristics for range-based search of the input domain $\mathit{\mathbb{D}_I}$ for identifying regions of equivalence. 
Recall that based on Definition~\ref{def:program}, input domain $\mathit{\mathbb{D}_I}$ is the Cartesian product of input parameter domains $D_{i_n}$, where $D_{i_n}$ denotes the possible values of input parameter $i_n ~(1 \leq n \leq N)$ and $N$ denotes the number of input parameters. For our range-search approach, we model each input parameter domain $D_{i_n}$ as a range of values represented as a pair $p_{i_n}= (\mathit{p_{i_n}.min, p_{i_n}.max})$ that denotes the minimum and maximum of the set of values in the domain (i.e., an interval). Then, $\mathit{\mathbb{D}_{I}}$ can be represented as a vector of pairs, $\overrightarrow{R_I} = [\mathit{p_{i_1}, p_{i_2},\ldots, p_{i_N}}]$. 

\paragraph{Relational range-search.} Algorithm \ref{proc:relational} is our first heuristic for range-based partial equivalence analysis. It divides the domains of the input parameters into subdomains and then checks for equivalence and total non-equivalence in those subdomains. The call to $\textsc{DivideRange}$ function divides the input domain of individual input parameters in equal halves and creates all possible combinations of the subdomains of all the input parameters. Suppose, two compared programs have 2 input variables $\overrightarrow{V_I} = [v_1, v_2]$ and the corresponding domains/ranges $\overrightarrow{R_I}= [\mathit{p_{i_1}, p_{i_2}}]$.

$\textsc{DivideRange}$ function first generates the middle points: $\mathit{p_{i_n}.mid =}$ $\mathit{p_{i_n}.min}$ 
$\mathit{+\lceil (p_{i_n}.max - p_{i_n}.min)/2 \rceil}$ of the input domains of the variables. Then, it returns a set of ranges $\mathit{R_{div}}$ consisting of vectors: \\
$\{[\mathit{(p_{i_1}.min, p_{i_1}.mid), (p_{i_2}.min, p_{i_2}.mid)}],$ \\
$[\mathit{(p_{i_1}.min, p_{i_1}.mid), (p_{i_2}.mid+1, p_{i_2}.max)}],$ \\
$[\mathit{(p_{i_1}.mid+1, p_{i_1}.max), (p_{i_2}.min, p_{i_2}.mid)}],$ \\
$[\mathit{(p_{i_1}.mid+1, p_{i_1}.max), (p_{i_2}.mid+1, p_{i_2}.max)}]\}$. \\
As shown here, with 2 input variables, a set of 4 different vectors will be generated. In general, $\textsc{DivideRange}$ generates $2^{N}$ different vectors for $N$ input parameters. 

\begin{algorithm}[t]
\caption{{\sc RelationalRangeSearch}($\overrightarrow{V_I}, \overrightarrow{R_I}, \mathit{depth}$)\\
Takes input variables $\overrightarrow{V_I}$, ranges of the input domain $\overrightarrow{R_I}$ and $\mathit{depth}$ as arguments, checks for equivalence and total non-equivalence recursively by dividing the input domains, and captures the equivalent ranges of input in the global variable $\mathit{R_{sol}}$ until either $\mathit{depth}$ reaches the predefined $\mathit{LIMIT}$ or reports equivalence/total non-equivalence in all divided ranges of the input domain.}
\label{proc:relational}
\begin{footnotesize}
\begin{algorithmic}[1]
\If{$\mathit{depth}=\mathit{LIMIT}$}
\State \Return
\EndIf
\State $\mathit{R_{div}} \leftarrow \textsc{DivideRange}(\overrightarrow{\mathit{R_I}})$
\For{$\overrightarrow{r_I}$ in $\mathit{R_{div}}$}
\State $\overrightarrow{\mathit{Min}} \leftarrow [\mathit{p'_{i_1}.min}, \mathit{p'_{i_2}.min},\ldots,\mathit{p'_{i_N}.min}]$ from $\overrightarrow{r_I}$
\State $\overrightarrow{\mathit{Max}} \leftarrow [\mathit{p'_{i_1}.max}, \mathit{p'_{i_2}.max},\ldots,\mathit{p'_{i_N}.max}]$ from $\overrightarrow{r_I}$
\If{$\neg \textsc{IsSat}( \neg (S_1 \land \mathit{\overrightarrow{V_I} \geq \overrightarrow{\mathit{Min}}} \land \mathit{\overrightarrow{V_I} \leq \overrightarrow{\mathit{Max}}} \Leftrightarrow S_2 \land \mathit{\overrightarrow{V_I} \geq \overrightarrow{\mathit{Min}}} \land \mathit{\overrightarrow{V_I} \leq \overrightarrow{\mathit{Max}}}))$}
\State $\mathit{R_{sol}} \leftarrow \mathit{R_{sol}} \cup \{\overrightarrow{r_I}\}$
\ElsIf{$\textsc{IsSat}(S_1 \land \mathit{\overrightarrow{V_I} \geq \overrightarrow{\mathit{Min}}} \land \mathit{\overrightarrow{V_I} \leq \overrightarrow{\mathit{Max}}} \land S_2 \land \mathit{\overrightarrow{V_I} \geq \overrightarrow{\mathit{Min}}} \land \mathit{\overrightarrow{V_I} \leq \overrightarrow{\mathit{Max}}})$}
\State {\sc RelationalRangeSearch}($\overrightarrow{V_I}, \overrightarrow{r_I}$, depth+1)
\EndIf
\EndFor
\State \Return
\end{algorithmic}
\end{footnotesize}
\end{algorithm}

   \begin{algorithm}[t]
\caption{{\sc Iterative}($ \overrightarrow{V_I},\overrightarrow{R_I}$)\\
Takes input variables $\overrightarrow{V_I}$, ranges of the continuous input domain $\overrightarrow{R_I}$ as arguments,  calls $\textsc{RelationalRangeSearch}$ for each pair of minimum, maximum values separately for each input variable and returns the corresponding $\mathit{R_{sol}}$'s as a list.}
\label{proc:iterative}
\begin{footnotesize}
\begin{algorithmic}[1]
 \State $\mathit{R_{solList}} \leftarrow \mathit{empty~list} $
\For{$p_{i_n}$ in $\overrightarrow{R_I}$  }
\State $\mathit{R_{sol}} \leftarrow \emptyset$
\State $\textsc{RelationalRangeSearch} ([{v_n}], [p_{i_n}], 0)$ 
 \\ \Comment{$v_n \in \overrightarrow{V_I} \mathit{~represents~individual~input~paramete}r$}
\State $\mathit{R_{solList}} \leftarrow \mathit{R_{solList}}+\mathit{R_{sol}}$
\EndFor
\State \Return $\mathit{R_{solList}}$
\end{algorithmic}
\end{footnotesize}
\end{algorithm} 
Each vector, $\mathit{\overrightarrow{r_I}}$, in the returned set $\mathit{R_{div}}$ contains a new range from the initial domain represented with a new pair $p'_{i_n}$ for each input variable $v_n$. In each iteration of the loop in Algorithm~\ref{proc:relational}, a vector $\overrightarrow{r_I}$ from $\mathit{R_{div}}$ is taken and $\mathit{\overrightarrow{\mathit{Min}}, \overrightarrow{\mathit{Max}}}$ are generated from the new minimum, maximum values of the new pairs $p'_{i_n}$ from $\overrightarrow{r_I}$. Then, new constraints are imposed on the the summaries $S_1, S_2$ of the programs for restricting the domains of all the input variables to the new ranges defined by the new minimum, maximum values, and the algorithm checks for equivalence (line 7) and total non-equivalence (line 9) in the new subdomains of the inputs. If the programs are found to be partially equivalent in the new subdomains, the algorithm is called recursively to divide the subdomains again for further equivalence analysis. It terminates either when $\mathit{depth}$ reaches the predefined $\mathit{LIMIT}$ or when it finds equivalence or total non-equivalence in all the divided ranges. Here, the upper limit of $\mathit{LIMIT}$ is 
$B = \log_2(|D_{i_n}|)$
(bits required to represent the size of the largest input domain $\mathit{D_{i_n}}$) and thus, the complexity of the algorithm in terms of the number of calls to the SAT solver is $\mathit{O((2^N)^{LIMIT})}$. Along the way, this algorithm collects the equivalent ranges in the global variable $\mathit{R_{sol}}$ where $\mathit{R_{sol}} = \{\overrightarrow{r_{I_1}}, \overrightarrow{r_{I_2}},\ldots, \overrightarrow{r_{I_T}}\}$ assuming there are $T$ different combinations of input subdomains for which the programs are equivalent.
Here, $\overrightarrow{r_{I_t}} = [p'_{i_1}, p'_{i_2},\ldots, p'_{i_N}]$ where $\overrightarrow{r_{I_t}}[n]=p'_{i_n}$ is a pair of minimum and maximum values that denotes a range of $n$-th input parameter from the $t$-th combination of input subdomains for which the programs are equivalent. Therefore, a lower bound on the size of equivalent domain can be computed as: 
\[
 \mathit{|\mathbb{D}_{eq}| \geq  \sum_{t=1}^{t= |R_{sol}|} \prod_{n=1}^{n=N} (\overrightarrow{r_{I_t}}[n].max - \overrightarrow{r_{I_t}}[n].min + 1)}
\]
Furthermore, the equivalence condition $F_{\textit eq}$ can be derived from the collected input ranges. However, this approach can become computationally expensive for multiple inputs as the number of solver calls increases exponentially with the number of input parameters.

\paragraph{Iterative range-search.} As relational range-search can blow up in terms of computation cost when the number of input parameters increases, we propose an iterative approach (Algorithm~\ref{proc:iterative}) as a more scalable solution. Rather than dividing the domains of all input parameters simultaneously, this approach  examines one parameter's domain at a time, leaving the domains of other input parameters unchanged. Then for each parameter, it calls Algorithm~\ref{proc:relational} using a vector with only the minimum and maximum values for that parameter's domain, disregarding the rest. This approach allows the algorithm to iteratively process each input parameter; in each iteration, $\mathit{R_{sol}}$ is updated with ranges of values for each input parameter where the programs exhibit equivalence. Finally, all equivalent ranges are stored in $\mathit{R_{solList}}$, where $\mathit{R_{solList}[n]}$ contains the range of values for the $n$-th input parameter's domain that result in program equivalence.
This iterative approach can reduce the number of calls to the SAT solver when dealing with multiple input parameters, and the corresponding complexity in terms of number of calls to the SAT solver becomes $O(\mathit{N} \times 2^\mathit{LIMIT})$. Now, for each variable, we can get the lower bound on equivalence: \\
 $\mathit{eq\_bound[n] = \sum_{t=1}^{t= |R_{solList}[n]|}(\overrightarrow{r_{I_t}}[0].max - \overrightarrow{r_{I_t}}[0].min + 1)}$ \\
and the upper bound on non-equivalence: \\
 $\mathit{neq\_bound[n] = |D_{i_n}| - eq\_bound[n]}$. \\ Therefore, the lower bound on equivalence can be computed as:
 \[
\mathit{|\mathbb{D}_{eq}| \geq  \sum_{n=1}^{n=N}  (\prod_{k=1}^{k=n-1} neq\_bound[k] \times eq\_bound[n] \times \prod_{j=n+1}^{j=N} |D_{i_j}|)}
\]

It is worth noting that although Algorithm~\ref{proc:iterative} is more scalable than Algorithm~\ref{proc:relational}, for cases involving multiple variables, it cannot capture relational equivalence conditions between variables. Additionally, Algorithms~\ref{proc:relational} and~\ref{proc:iterative} both may fail to yield results if none of the subdomains demonstrate either equivalence or total non-equivalence within the predefined depth limit, $\mathit{LIMIT}$.

\begin{algorithm}[t]
\caption{{\sc Priority}($v_n, p_{i_n}$)\\
Takes input variable $v_n$ and a range $p_{i_n}$ of the corresponding input variable as arguments, checks for equivalence and total non-equivalence by dividing the continuous domain of input variable and captures the equivalent ranges in global variable $\mathit{R_{sol}}$.}
\label{proc:priority}
\begin{footnotesize}
\begin{algorithmic}[1]
\State $\mathit{Min}, \mathit{Max} \leftarrow \mathit{p_{i_n}.min}, \mathit{p_{i_n}.max}$ 
\If{$\neg \textsc{IsSat}( \neg (S_1 \land \mathit{v_n \geq \mathit{Min}} \land \mathit{v_n \leq \mathit{Max}}  \Leftrightarrow S2 \land \mathit{v_n \geq \mathit{Min}} \land \mathit{v_n \leq \mathit{Max}}))$}
\State $r_{i_n} \leftarrow \textsc{expandDomainBoundary}(p_{i_n})$
\State $\mathit{R_{sol}} \leftarrow \mathit{R_{sol}} \cup \{r_{i_n}\}$
\State \Return
\EndIf
\If{$\mathit{Max} \leq \mathit{Min} \vee \neg \textsc{IsSat}(S_1 \land \mathit{v_n  \geq \mathit{Min}} \land \mathit{v_n \leq \mathit{Max}} \land S_2 \land \mathit{v_n  \geq \mathit{Min}} \land \mathit{v_n  \leq \mathit{Max}})$}
\State \Return
\EndIf
\State $ p'_{i_n} \leftarrow \textsc{PrioritizedDivideRange}(p_{i_n})$
\State \textsc{Priority}($v_n,  p'_{i_n}$)
\end{algorithmic}
\end{footnotesize}
\end{algorithm}
\begin{algorithm}[t]
\caption{{\sc IterativePriority}($ \overrightarrow{V_I},\overrightarrow{R_I}$)\\
Takes input variables $\overrightarrow{V_I}$, ranges of the continuous input domain $\overrightarrow{R_I}$ as arguments,  calls $\textsc{Priority}$ for each pair of minimum, maximum values separately for each input variable and returns the corresponding $\mathit{R_{sol}}$'s as a list.}
\label{proc:iterativePriority}
\begin{footnotesize}
\begin{algorithmic}[1]
 \State $\mathit{R_{solList}} \leftarrow \mathit{an~empty~list} $
\For{$p_{i_n}$ in $\overrightarrow{R_I}$  }
\State $\mathit{R_{sol}} \leftarrow \emptyset$
\State $\textsc{Priority} ({v_n}, (1, p_{i_n}.max))$
\State $\textsc{Priority} ({v_n}, (p_{i_n}.min, -1))$
\State $\textsc{Priority} ({v_n}, (0, 0))$
\State $\mathit{R_{solList} \leftarrow R_{solList} + R_{sol}}$
\EndFor
\State \Return $\mathit{R_{solList}}$
\end{algorithmic}
\end{footnotesize}
\end{algorithm}

\paragraph{Iterative priority range-search.} Algorithm~\ref{proc:iterative} divides the input domain recursively with each divided range having equal priority. Alternatively, we propose iterative priority range-search (Algorithm~\ref{proc:iterativePriority}, which relies on Algorithm~\ref{proc:priority}) which divides input domain with priority. Initially, this algorithm divides the domain of an input variable into three disjoint partitions: \textit{negative}, $0$, and \textit{positive} values. Then Algorithm~\ref{proc:priority} is called with these new partitions of the input domains for each input variable $v_n$. Algorithm~\ref{proc:priority} checks for equivalence of the programs within the given partition. If they are not equivalent/totally non-equivalent, it then calls $\textsc{PrioritizedDivideRange}$. It only considers the half of the partition which is closer to the value $0$, discarding the other half. So, for $\mathit{(1, p_{i_n}.max)}$, the maximum value is changed to $\lceil (p_{i_n}.max / 2) \rceil$ and for $\mathit{(p_{i_n}.min, -1)}$, the minimum is changed to $\lceil (p_{i_n}.min / 2) \rceil$. In these two cases, the priority of division is focused towards the value $0$ and this terminates when $\mathit{Min}$ matches or exceeds $\mathit{Max}$. The algorithm may terminate early, if it finds equivalence/total non-equivalence in the divided subdomain. Additionally, it attempts to push the boundary of that subdomain towards the previously discarded half of that partition by calling \textsc{ExpandDomainBoundary} which then checks for equivalence there. Thus, it can get an accurate continuous subdomain of equivalence centering around $0$ whenever applicable.
For the partition $(0, 0)$, the algorithm just checks for equivalence at $0$ and terminates. The intuition behind this heuristic is that programs are more likely to be equivalent for the smaller absolute values in the input domain.

This priority search iteratively shrinks the ranges of both \textit{positive} and \textit{negative} partitions towards the center i.e., $0$ and so, the complexity with respect to the number of calls to the SAT solver becomes $O(\mathit{N} \times \log 2^B) = O(\mathit{N} \times \mathit{B})$.
As this only changes the way of dividing the input space, the lower bound on equivalence can be computed using the same formula used for the iterative range-search.

\begin{algorithm}[t]
\caption{{\sc CombinedRangeSearch}($ \overrightarrow{V_I},\overrightarrow{R_I}$)\\
Takes input variables $\overrightarrow{V_I}$, ranges of the continuous input domain $\overrightarrow{R_I}$ as arguments,  calls the iterative Algorithms~\ref{proc:iterative}, ~\ref{proc:iterativePriority} and saves the best lower bound on equivalence result in $\mathit{R_{solFinal}}$ .}
\label{proc:combinedRangeSearch}
\begin{footnotesize}
\begin{algorithmic}[1]
 \State $\mathit{R_{solList1}} \leftarrow \textsc{Iterative} (\overrightarrow{V_I}, \overrightarrow{R_I}) $
  \State $\mathit{R_{solList2}} \leftarrow \textsc{IterativePriority} (\overrightarrow{V_I}, \overrightarrow{R_I}) $
  \State ${\mathit{R_{solFinal}}} \leftarrow \mathit{an~empty~list}$
\For{$k$ in $N: N = \mathit{number~of~parameters}$}
\If{$\mathit{lower~bound~on~R_{solList1}}[k] \geq \mathit{R_{solList2}}[k]$}
\State ${\mathit{R_{solFinal}} \leftarrow \mathit{R_{solFinal}} + \mathit{R_{solList1}[k]}}$
\Else
\State ${\mathit{R_{solFinal}} \leftarrow \mathit{R_{solFinal}} + \mathit{R_{solList2}[k]}}$
\EndIf
\EndFor
\State \Return $\mathit{R_{solFinal}}$
\end{algorithmic}
\end{footnotesize}
\end{algorithm}

\paragraph{Combined range-search.} Algorithm~\ref{proc:combinedRangeSearch} combines both of our iterative approaches, Algorithms~\ref{proc:iterative} and~\ref{proc:iterativePriority}, for providing a better partial equivalence analysis by choosing the best result for each input.



\subsection{Partial Equivalence Analysis with Model Counting}
\label{sec:baselines}

To establish a baseline for the range-search based partial equivalence analysis, we introduce two alternative approaches. The first one employs enumerative model counting, while the second utilizes formula projection combined with model counting techniques.


\paragraph{Enumerative Model Counting.} Enumerative model counting takes a formula $F$ as input and finds the satisfying solutions for $F$ iteratively by calling an SMT solver, projected only on the input variables $\overrightarrow{V_I}$. In each iteration, the formula $F$ gets updated by adding a new constraint to avoid getting a duplicate solution in next iteration. Given a time limit $\mathit{TL}$, this approach can lead to 3 possible cases.

\textbf{Case 1:} When all possible elements of equivalent input domain $\mathbb{D}_{\textit eq}$ (iterating solutions of $(S_1 \land S_2)$) are obtained within the given time limit $\mathit{TL}$. 

\textbf{Case 2:} When all possible elements of non-equivalent input domain $\mathbb{D}_{\textit neq}$ (iterating solutions of $\neg (S_1 \Leftrightarrow S_2)$) are obtained within the given time limit $\mathit{TL}$, and, therefore, $\mathbb{D}_{\textit eq} = \mathbb{D}_{I} - \mathbb{D}_{\textit neq}$ can also be found. 

\textbf{Case 3:} Neither case 1 nor case 2 holds. So, this gives us a subset of $\mathbb{D}_{\textit eq}$ and a subset of $\mathbb{D}_{\textit neq}$ within the given time limit $\mathit{TL}$, and we get a lower bound on equivalence. 


The complexity of this approach with respect to the number of calls to the solver is equal to the number of satisfiable solutions of the formula $F$. In the worst case, it can result in $O(2^{BN})$ calls.
So, this enumerative approach can be expensive in getting the exact solutions under limited resources if the number of satisfiable solutions is significantly high.

\paragraph{Projection and Model Counting.} 
To derive input conditions under which programs $P_1$ and $P_2$ behave equivalently, we existentially quantify all output variables in the formula $S_1 \Leftrightarrow S_2$, i.e., $F \leftarrow \exists o_1, \ldots, \exists o_M. (S_1 \Leftrightarrow S_2)$, and apply SMT-based simplification to project this formula onto input variables: $F_{\textit{eq}} \leftarrow \textsc{Simplify}(F)$. We then use model counting tools~\cite{searchMC,ganak,qCoral,ABC} to compute the equivalence percentage by counting satisfying solutions to $F_{\textit{eq}}$. However, both quantifier elimination and model counting are computationally expensive, potentially limiting scalability.

\section{Patch Dataset}
\label{sec:patchBench}

We extracted a set of patches targeting numeric error CVEs from 2007 to 2019~\cite{cweNumeric,classifyPatches} across three large open-source projects: Linux~\cite{linux}, Qemu~\cite{qemu}, and FFmpeg~\cite{ffmpeg}. For each vulnerability, we constructed a paired code artifact consisting of the original vulnerable version and its corresponding patched version. For example, from Listing~\ref{lst:ffmpeg_189_0859}, we derived the original and patched code segments shown in Listing~\ref{lst:originalAndPatchFFmpeg}. The patched version reflects the changes applied relative to the original code.

\begin{lstlisting}[caption=The original and patched versions from Listing~\ref{lst:ffmpeg_189_0859},label=lst:originalAndPatchFFmpeg, language=C, xleftmargin=7.0ex,  escapeinside={(*@}{@*)},numbers=none,]
(*@\textbf{\textcolor{red}{Original Version}} @*) 
static int add_doubles_metadata_Snippet(int count){
    if (count >= INT_MAX / sizeof(int64_t)){
        return AVERROR_INVALIDDATA;}
    return count * sizeof(int64_t);}

(*@\textbf{\textcolor{green!50!black}{Patched Version}} @*)
static int add_doubles_metadata_Snippet(int count){
    if (count >= INT_MAX / sizeof(int64_t) || count <= 0)
        return AVERROR_INVALIDDATA;
    return count * sizeof(int64_t);}
    
\end{lstlisting}

In addition, we collected patches from the Juliet Test Suite~\cite{boland2012juliet}, which consists of programs annotated with known Common Weakness Enumerations (CWEs). The benchmark distinguishes between good and bad patches~\cite{black2018juliet}: bad patches eliminate the vulnerability by overly restrictive means, such as hard-coding safe values, whereas good patches remediate the vulnerability by appropriately conditioning on the inputs that trigger it. We collected both good and bad patches along with their corresponding original (vulnerable) programs. Furthermore, we selected a subset of the real-world patches collected from Linux, Qemu, and FFmpeg and used a large language model (LLM), GPT-4o~\cite{openai_gpt4o}, to synthesize bad patches for each selected case. These LLM-generated bad patches were constructed to follow the same over-restrictive patterns observed in the bad patches of the Juliet Test Suite, such as fixing the vulnerability via hard-coded values or excessive input restriction rather than input-sensitive checks. In this experimental setup, we treat the original developer generated patches as good patches and the LLM-generated patches as their corresponding bad patches.

\section{Evaluation}
\label{sec:implmentation}

\paragraph{Implementation.} We implemented our partial equivalence analysis for C programs using SMT bit-vector theory and compiled binaries to obtain precise results. The angr~\cite{angr} framework was used to implement Algorithm~\ref{proc:classifier}, where \textsc{Summarize} generates bit-vector encoded symbolic summaries. All range-search algorithms rely on satisfiability checking with Z3 solver. Algorithms~\ref{proc:relational} and~\ref{proc:iterative} use a depth limit ($\mathit{LIMIT}$) set empirically to 8 for single-variable and 4 for multi-variable programs. The enumerative method uses Z3 to find satisfying solutions, while the projection-based method leverages Z3’s quantifier elimination. For model counting, we use the probabilistic counters SearchMC~\cite{searchMC} and Ganak~\cite{ganak}, the compositional statistical counter qCoral~\cite{qCoral} which applies interval constraint propagation, and the exact automata-based counter ABC~\cite{ABC}, which supports linear and string constraints. Our implementation and dataset has been made available in a git repository~\cite{ourAnonRepo}. 

\label{sec:experiment}

\paragraph{Experimental Setup.} We evaluated our patch impact analysis on PatchBench with 90 program pairs from Linux (48), Qemu (16), and FFmpeg (26), and JulietBench with 50 programs containing good and bad patches, totaling 100 pairs. Of these, 11 programs are taken directly from the Juliet Test Suite and cover numeric CWEs 190, 191, and 369; the remaining JulietBench programs are derived from PatchBench, with bad patches generated by an LLM. We also analyzed 47 C programs from EqBench~\cite{eqBench}, finding partial equivalence in 28 cases. The number of lines of codes (LOC) of the programs in our dataset ranges between 8 to 64 with an average LOC of 24.76, 23.74, and 16.50, respectively for PatchBench, JulietBench and EqBench dataset. Methods are referred to as Enumeration (enumerative approach) and by the model counters in projection-based approaches. Experiments ran on a 13th Gen Intel Core i9-13900K, 192 GB RAM, Ubuntu 22.04.4 LTS, angr v9.2.23, Python 3.8.10, and Z3 v4.10.2.

\paragraph{Research Questions.} Our experimental evaluation targets the following research questions: \\
\textbf{RQ1.} Is our approach effective in assessing patch impact given two versions of a program? \\
\textbf{RQ2.} How does Relational (Algorithm~\ref{proc:relational}) compare to Combined (Algorithm~\ref{proc:combinedRangeSearch}) RangeSearch in performing partial equivalence analysis?\\
\textbf{RQ3.} How effective and efficient is our approach (Algorithm~\ref{proc:combinedRangeSearch}) in analyzing quantitative partial equivalence compared to model counting approaches?

\textbf{{\em RQ1 : Patch Impact Analysis.}} 
\begin{figure}[t]
    \begin{minipage}[b]{0.5\textwidth}
        \includegraphics[width=\textwidth]{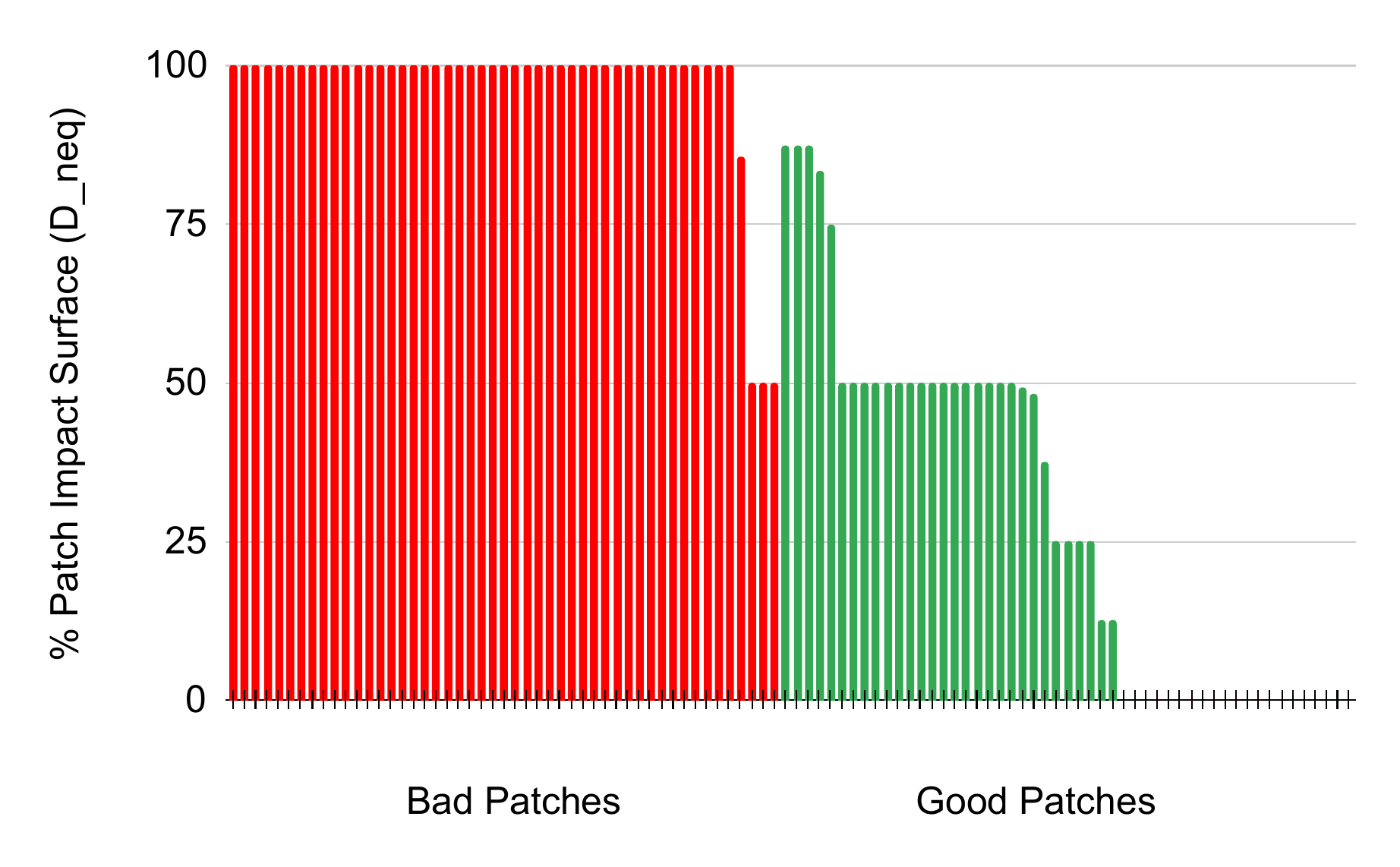}  
    \end{minipage}
    \hfill
    \begin{minipage}[b]{0.5\textwidth}
        \includegraphics[width=\textwidth]{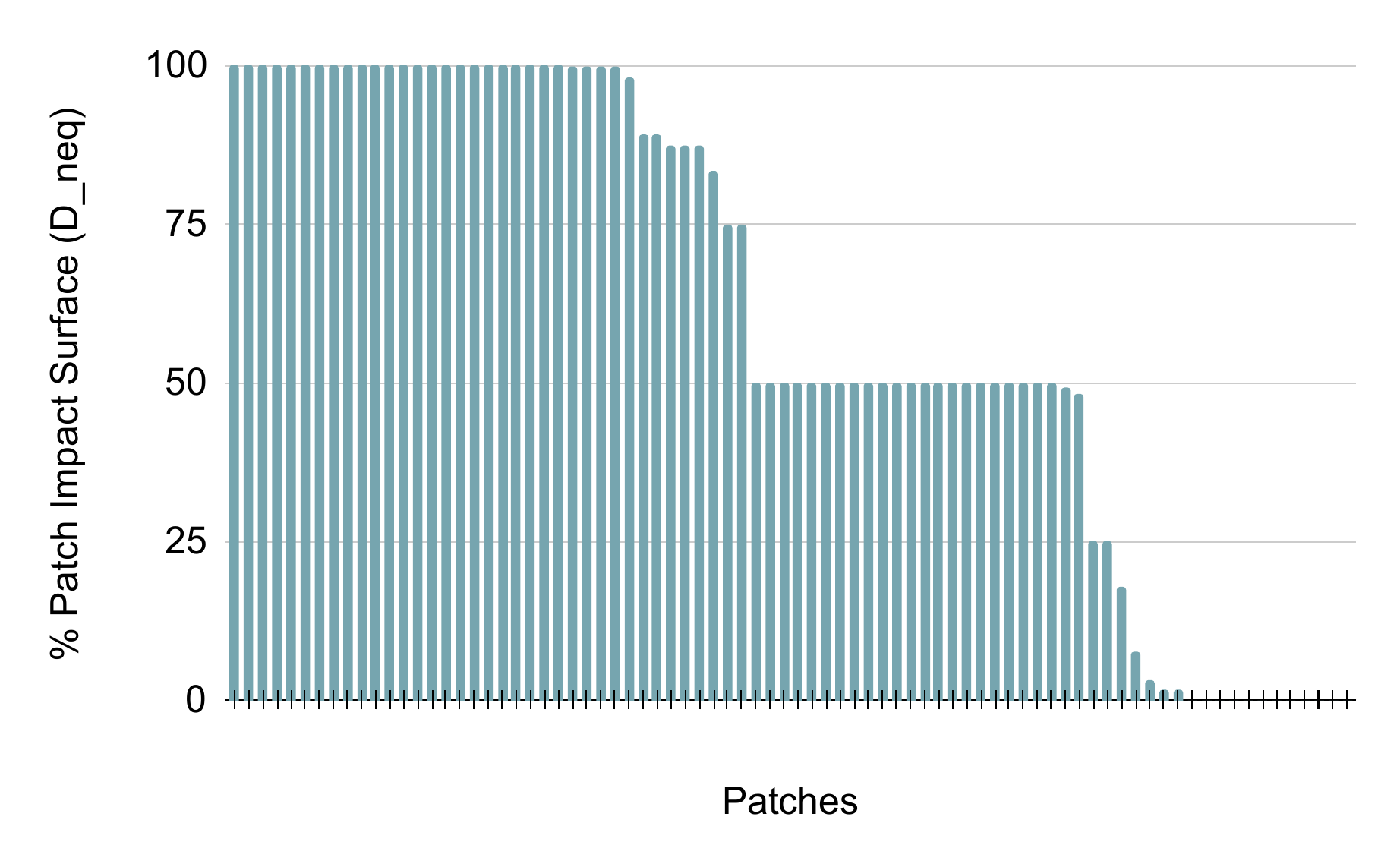}  
    \end{minipage}
    \caption{Patch impact analysis results showing the size of the patch impact surface (as the percentage of the input domain) (Left) for good \& bad patches in JulietBench Dataset  and (Right) for different patches in PatchBench.}
    \label{fig:patchimpactanalysis}
   \vspace*{-0.2in}
\end{figure}
\begin{table*}[t]
\scriptsize
\captionof{table}{Equivalence Conditions Computed with Partial Equivalence Analysis}
\label{tab:partialEqAnalysis}
\begin{tabular}{|l|l|l|l|}
\hline
\textbf{Benchmark}   & \textbf{Row}        & \textbf{Program Pair Source}                & \textbf{Equivalence Condition}                                                    \\ \hline 
  &1 & Qemu:CWE-369:CVE\_2014\_8e   & $0 < x < 8388608$                                                            \\ \cline{2-4} 
                             &2 & Qemu:CWE-369:CVE\_2014\_93   & $0 \leq x \leq 2147483647$                                                   \\ \cline{2-4} 
                               
                             &3 & FFmpeg:CWE-189:CVE\_2013\_39 & $0 \leq x \leq 536870911$                                                        \\ \cline{2-4} 
                             &4 & FFmpeg:CWE-190:CVE\_2016\_8a & $0 \leq x \leq 1073741823$ 
                             \\ \cline{2-4} 
                             &5 & Linux:CWE-189:CVE\_2010\_fd  & $2 \leq x \leq 2147483647$                                                   \\ \cline{2-4} 
                             &6 & Linux:CWE-189:CVE\_2012\_44   & $536870911 < x \leq 4294967295$ \\ \cline{2-4}
                             &7 & Linux:CWE-189:CVE\_2012\_ed  & $0 \leq x \leq 76695844$
                                                                            \\ \cline{2-4} 
                              &8 &  Linux:CWE-190:CVE\_2017\_0f   & $(0 \leq x \leq 4294967280) \wedge (0 \leq y \leq 4294967295)$                                           \\ \cline{2-4} 
\multirow{-9}{*}{PatchBench}     &9 & Linux:CWE-369:CVE\_2017\_49 & $(x > 25) \wedge (x \leq 65535)$                               \\ \hline \hline
                            &10   & Bad patch multiply:CWE-190                      & $x == 1$   
\\ \cline{2-4}
 \multirow{-3}{*}{JulietBench}               &11  & Good patch multiply:CWE-190                       & $x < 1073741823$
      \\ \hline \hline
                         &12     & oneN2               & $x \neq 2147483648$                                                                              \\ \cline{2-4} 
  \multirow{-3}{*}{EqBench}  &13 & dart                            & $(-1290 \leq x \leq 1290) \vee (y \neq 10 \wedge y \neq 20)$              
 
 \\ \hline

                           
\end{tabular}
\vspace*{-0.2in}
\end{table*}
As defined in the Juliet Test Suite, bad patches eliminate vulnerabilities by replacing bad inputs with hardcoded safe values. As a result, bad patches are expected to have larger patch impact surfaces compared to good patches, which preserve functionality by focusing only on vulnerable inputs. Fig.~\ref{fig:patchimpactanalysis} (left) illustrates this distinction by showing the patch impact surfaces of good patches (green) and bad patches (red) separately. On average, bad patches exhibit a patch impact surface of 96.65\%, whereas good patches affect only 29.03\% of the input domain. These results demonstrate that our patch impact analysis quantitatively captures the behavioral differences and can identify patches that warrant additional scrutiny due to excessive impact on the input domain.


Fig.~\ref{fig:patchimpactanalysis} (right) presents the patch impact surface distribution for the PatchBench dataset. We observe that 36.25\% of the patches are non-equivalent for more than 90\% of the input domain, while 20\% are non-equivalent for less than 10\% of the input domain. This wide variance indicates that real-world patches differ substantially in terms of their patch impact surface. Overall, our patch impact analysis exposes divergent behavioral effects of patches and provides a quantitative measure for comparing their impact on input domain.

Table~\ref{tab:partialEqAnalysis} shows the input equivalence conditions on a subset of the partially equivalent programs from our dataset. The equivalence condition helps us understand for which inputs given two programs are equivalent/non-equivalent. For example, in CWE-369: CVE\_2014\_93 patch from Qemu (Table~\ref{tab:partialEqAnalysis} row 2), a datatype change from $\mathit{int}$ to $\mathit{unsigned~int}$ results in equivalence only within the input range $0 \leq x \leq 2147483647$, accounting for 49.99\% of the input domain. In rows 10 and 11, we can find equivalence conditions for good and bad patches where the bad patch is equivalent only for one hardcoded fixed value and on the contrary, good patch is equivalent for $x < 1073741823$.

Additionally, even though EqBench is designed for equivalence analysis, not directly for patch assessment, our analysis identified 5 mislabeled “equivalent” cases in EqBench. In these cases, integer overflows render the programs non-equivalent (one example is discussed in Appendix~\ref{appendix}). Our approach reports both the input conditions and the extent of non-equivalence for these programs (Table~\ref{tab:partialEqAnalysis}, Rows 12–13). \textbf{These results demonstrate that our patch impact analysis can effectively quantify behavioral divergence and determine equivalence boundaries between two program versions.}

\begin{figure}[t]
     \begin{minipage}[b]{0.4\textwidth}
\includegraphics[width=\textwidth]{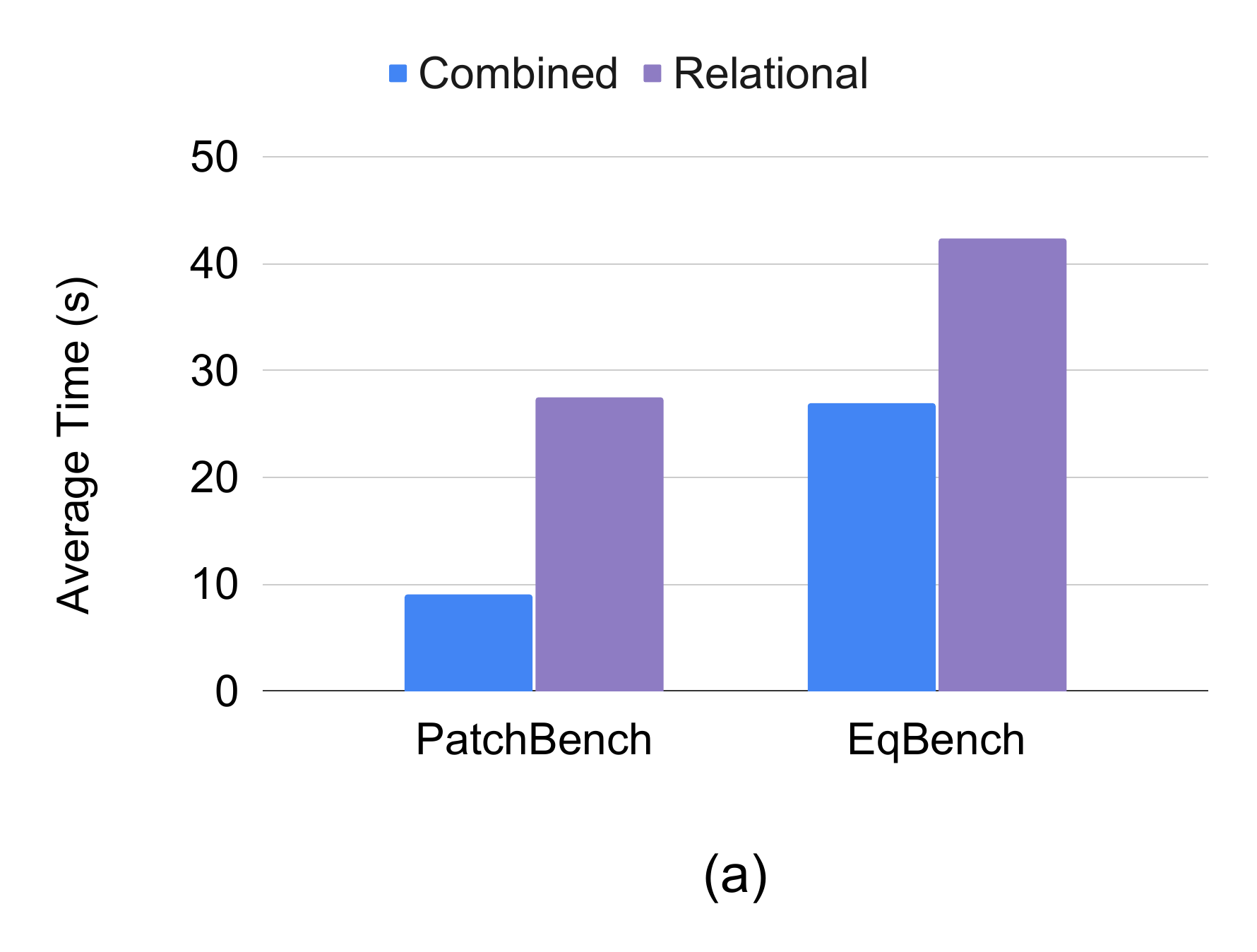}  
    \end{minipage}
    \hfill
    \begin{minipage}[b]{0.5\textwidth}
        \includegraphics[width=\textwidth]{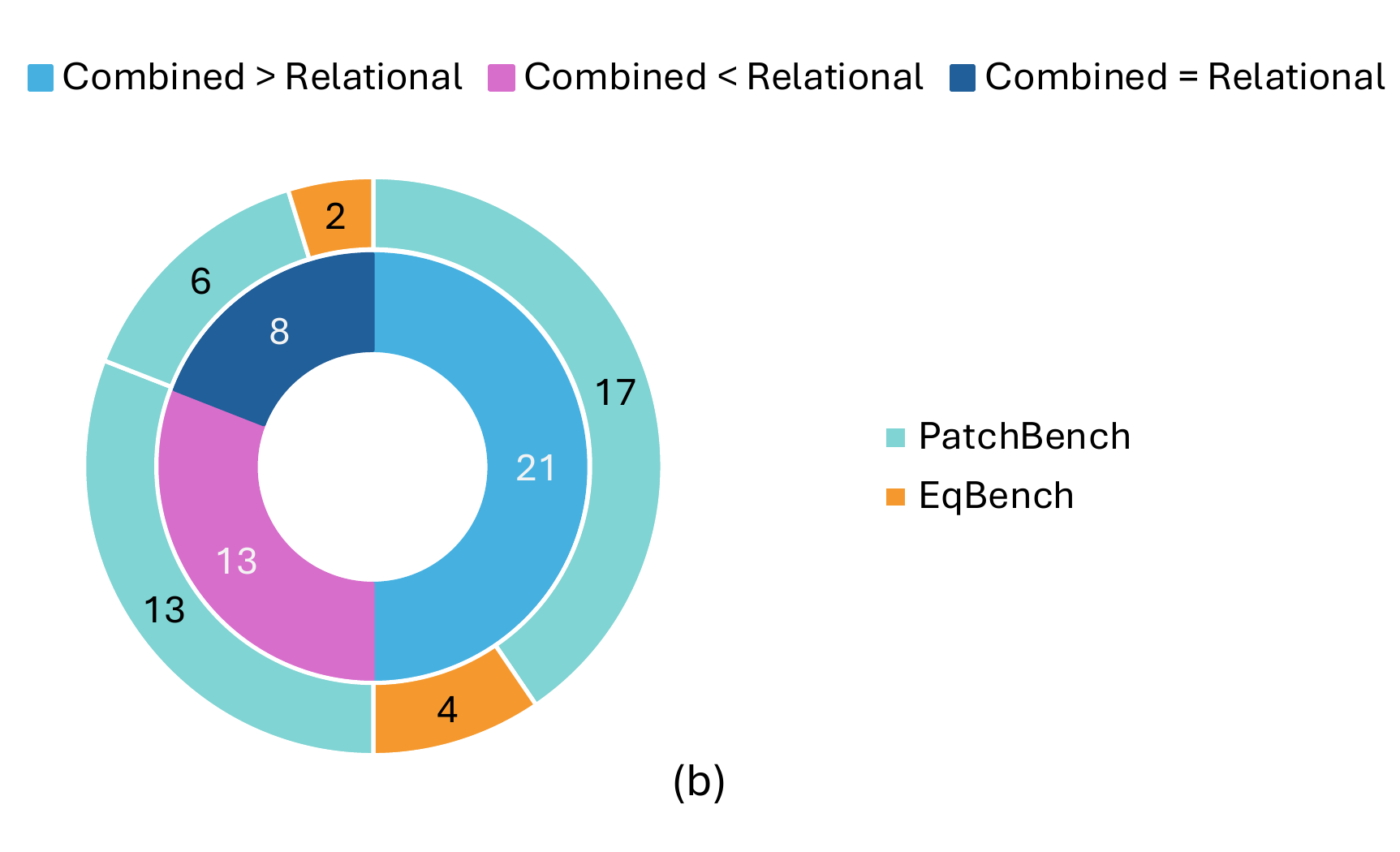}  
    \end{minipage}
     \caption{Comparison of Combined and Relational Range-Search approaches:  (a) Average execution times for PatchBench and EqBench. (b) The inner circle shows cases where Combined outperforms Relational (Combined > Relational), where Relational outperforms Combined (Combined < Relational), and where both perform equally (Combined = Relational) in terms of the equivalence bound they compute. The outer circle denotes distributions of cases between PatchBench and EqBench.}
    \label{fig:relationalVsCombined}
\vspace*{-16pt}   
\end{figure}

\textbf{{\em RQ2: Comparison of Relational (Algorithm~\ref{proc:relational}) and Combined (Algorithm~\ref{proc:combinedRangeSearch}) RangeSearch.}}
Recall that \textsc{RelationalRangeSearch} (Algorithm~\ref{proc:relational}) considers all the input parameters at once while dividing them into ranges, unlike the iterative one. To compare \textsc{RelationalRangeSearch} (Algorithm~\ref{proc:relational}) with  \textsc{CombinedRangeSearch} (Algorithm~\ref{proc:combinedRangeSearch}), we focus on programs with more than one input parameters. Out of the 118 program pairs from both datasets, there are 42 programs with more than 1 input parameter (ranging from 2 to 5 parameters). Fig.~\ref{fig:relationalVsCombined}(a) shows the timing performance of the two algorithms on these programs. For both datasets, the Relational range-search algorithm has a higher time overhead than the Combined algorithm, which is expected as the time complexity increases exponentially with the number of input variables for the Relational algorithm whereas it increases linearly for the Combined algorithm. 

Fig.~\ref{fig:relationalVsCombined}(b) shows the performance comparison with respect to finding the equivalence condition and corresponding percentage for both approaches. In 21 of the cases, Combined range-search has provided a better equivalence bound than the Relational one, whereas in 13 of the cases, Relational one does better. To showcase an example where Relational approach performs better than Combined, consider the patched version of Linux:CWE\_190:CVE\_2018\_fb. An additional check $(\mbox{\tt nr\_wake} < 0 \vee \mbox{\tt nr\_requeue} < 0)$ is added by the patch and thus two versions become equivalent only when $(\mbox{\tt nr\_wake} \geq 0 \wedge \mbox{\tt nr\_requeue} \geq 0)$ where {\tt nr\_wake}, {\tt nr\_requeue} are input parameters of $\mathit{signed}$ 32-bit integer datatype. As there is a relational condition between the two input parameters for the patch versions to be equivalent,  our Combined algorithm fails to capture the equivalence condition precisely, whereas the Relational one succeeds in capturing this condition. For rest of the 8 cases, both approaches report the same equivalence bounds. Note that there are no cases in EqBench where the Relational one has performed better than Combined which means that relational constraints involving multiple input parameters are absent in that dataset. \textbf{To summarize, if there is no relational constraint involving multiple input parameters, Combined performs better than the Relational range-search.} 

\textbf{{\em RQ3 : Effectiveness of RangeSearch (Algorithm~\ref{proc:combinedRangeSearch}) compared to model counting approaches.}} 
\label{subsubsection:rq2}
\begin{figure}[t]
    \centering
    \begin{minipage}[b]{0.49\textwidth}
        \centering
        \includegraphics[width=\textwidth]{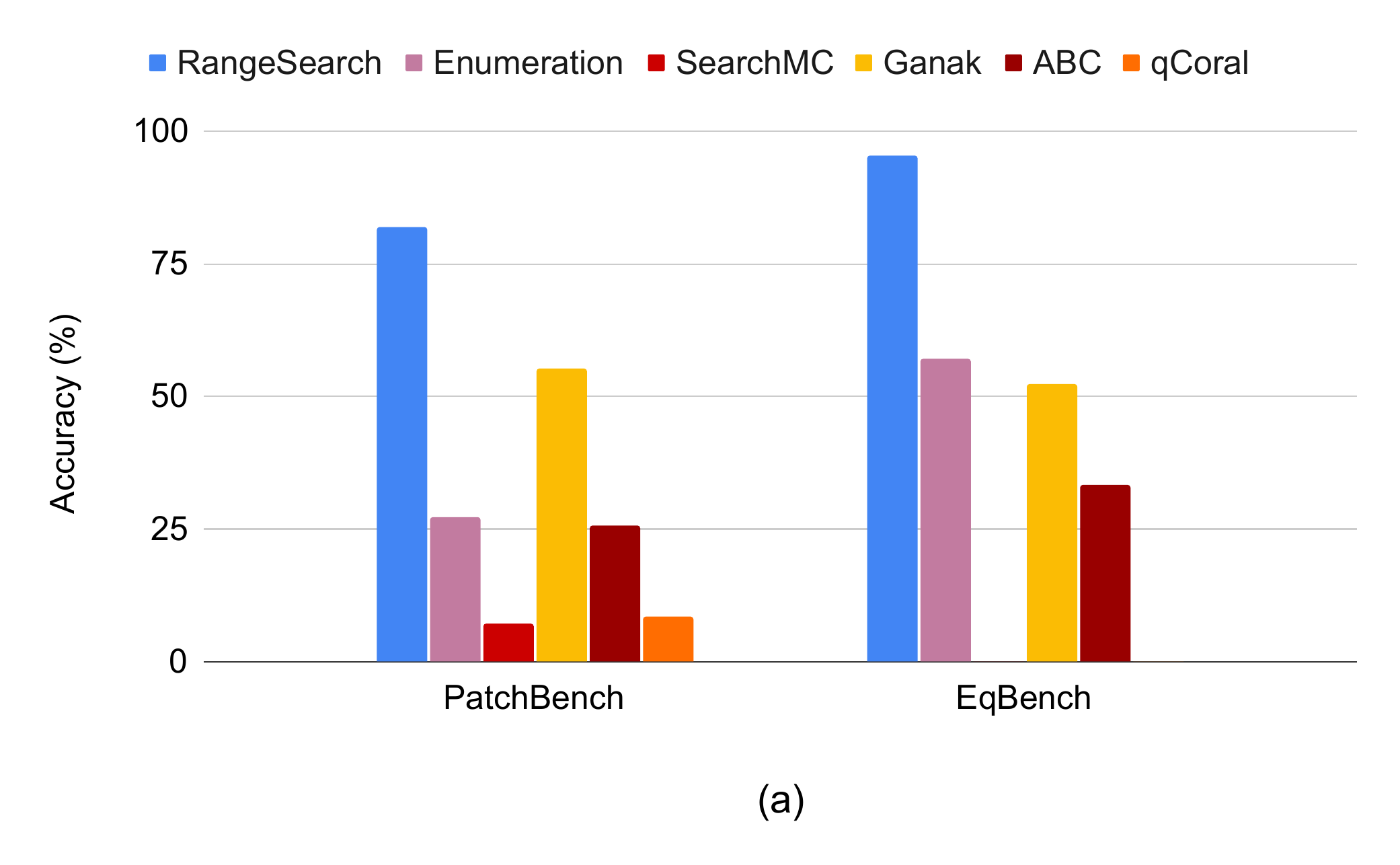}  
    \end{minipage}
    \hfill
    \begin{minipage}[b]{0.49\textwidth}
        \centering
        \includegraphics[width=\textwidth]{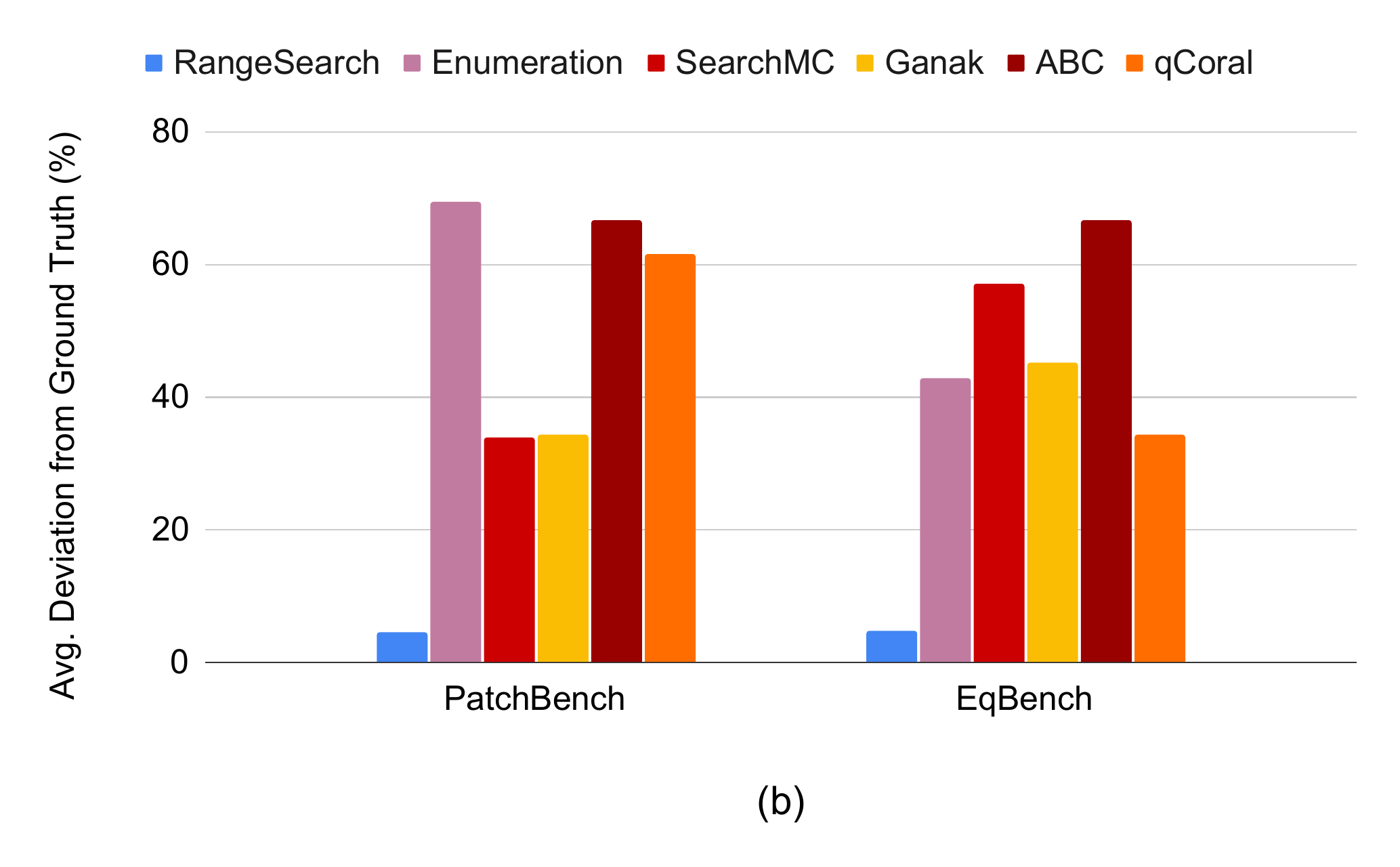}  
    \end{minipage}
    \caption{Performance comparison of different approaches on equivalence bound: (a) Accuracy of achieving exact result (higher is better). (b) Difference between ground truth and the reported equivalence bound (lower is better).}
    \label{fig:exactResultComp}
   \vspace*{-0.1in}
\end{figure}
\begin{figure}[t]
    \centering
\includegraphics[width=.49\textwidth]
    {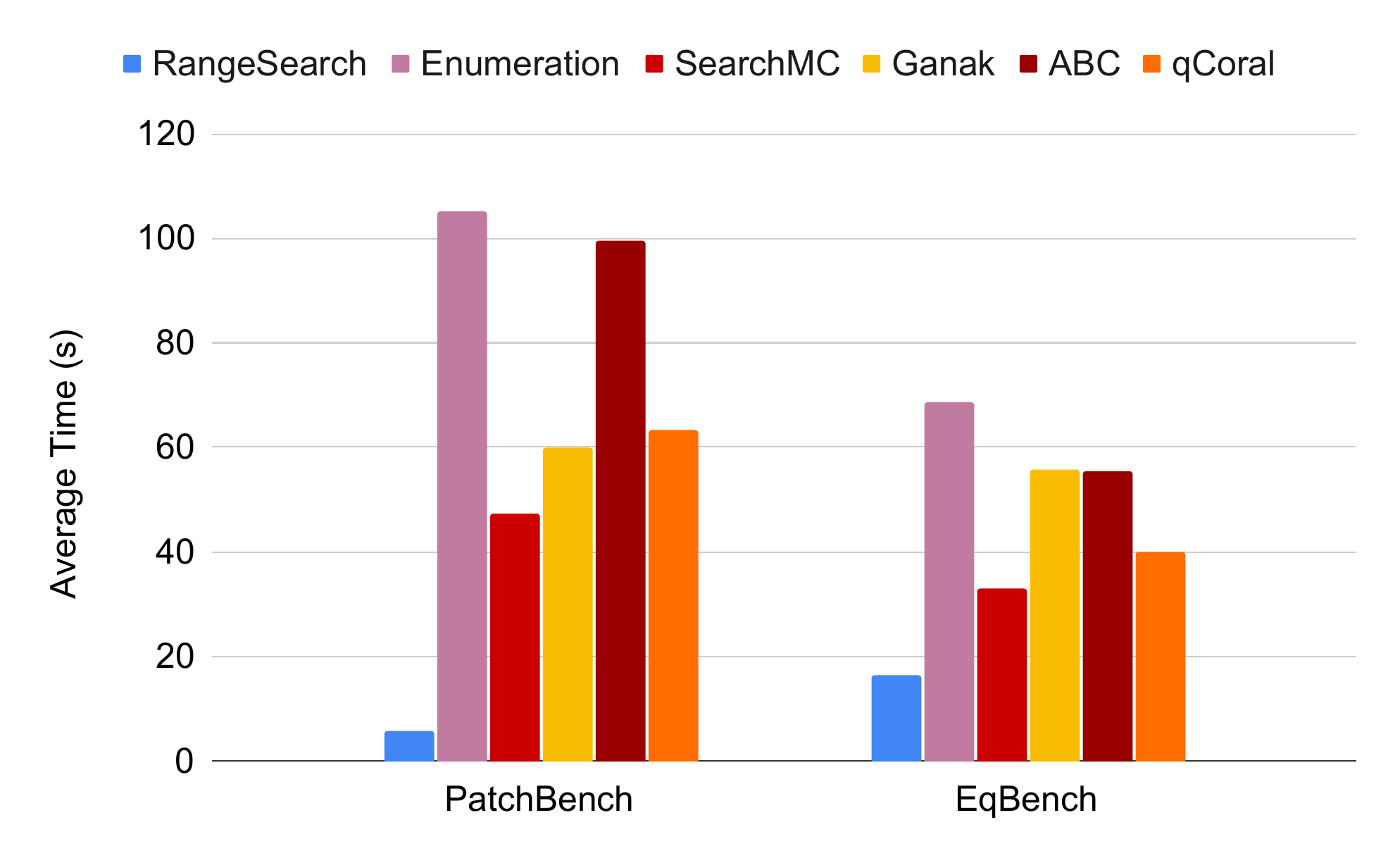}
    \caption{Average analysis time for different approaches.}
    \label{fig:timeComp}
        \vspace*{-0.2in}
\end{figure}
To evaluate our approach, we first analyzed partial equivalence across the dataset. Enumeration performs a complete search over the input domain, typically producing equivalence conditions with an average of 14,871 enumerated solutions within a 120-second time limit. It is more efficient when partial equivalence involves fewer input solutions. Projection-based model counting methods rely on projection to determine equivalence conditions and model counting to calculate equivalence percentages. Among 118 cases, projection failed in 22\% within the time limit (120-second). Effectiveness for the remaining cases depends on the model counters’ performance. Our proposed method, RangeSearch (henceforth referring to Algorithm~\ref{proc:combinedRangeSearch}), searches within specified ranges, providing a sound lower bound on equivalence. Across all 118 cases, RangeSearch yields equivalence conditions and percentages in 81.4\% of cases, compared to 61.7\%, 50.5\%, 18.6\%, and 39.5\% for SearchMC, Ganak, ABC, and qCoral, respectively.


We then evaluated exact equivalence bounds on 65 cases (44 from PatchBench, 21 from EqBench) with manually determined ground truth. Fig.~\ref{fig:exactResultComp}(a) illustrates the performance of different approaches in providing exact equivalence results. RangeSearch outperforms others, achieving 81.8\% accuracy on PatchBench and 95.2\% on EqBench, improving over Ganak (the second-best approach for PatchBench) by 26.6\% and Enumeration (the second-best for EqBench) by 38.1\%.


Next, we examine how closely each approach approximates the ground truth when they are unable to produce an exact result, as illustrated in Fig.~\ref{fig:exactResultComp}(b). Across both datasets, RangeSearch achieves the closest approximation to the ground truth on average compared to other methods. Although our approach's difference from the ground truth is not zero, it is significantly lower than that of other methods, approximately 29.4\% lower than the second-best approach on both datasets. 


Average execution times are reported in Fig.~\ref{fig:timeComp}. Enumeration is slowest due to exhaustive search, and ABC often times out on constraints with large constants. RangeSearch runs roughly 8× faster than SearchMC (the second-best) on PatchBench and 2× faster on EqBench. \textbf{Overall, these results demonstrate that RangeSearch outperforms other approaches in partial equivalence analysis.} 



\textbf{{\em Threats to Validity.}}  
Threats to internal validity pertain to the correctness of our RangeSearch implementation and the execution of our experiments. A comparison of our results with ground truth indicates that our approach functions as intended. One significant hyperparameter was the time limit, which primarily affects the enumeration approach but does not alter the overall results presented. Regarding external validity, concerns may arise from the selection of programs used in our experiments. We evaluated our approach on different distinct datasets: one focused on equivalence analysis and the others derived from patches.
Consequently, we believe our approach is generalizable for analyzing partial equivalence for numeric input domains.

\textbf{{\em Limitations.}} Our quantitative partial equivalence analysis relies on differential symbolic execution for summary generation (Algorithm~\ref{proc:classifier}) and therefore inherits the limitations of symbolic execution. If Algorithm~\ref{proc:classifier} fails to generate symbolic summaries or determine equivalence~\cite{dse}, the quantitative analysis cannot proceed.
There is ongoing research on efficient symbolic summary generation based on common code abstraction~\cite{dse,ardiff,pasda}, under-constrained symbolic execution~\cite{ramos2015under}, taint analysis based slicing~\cite{shafiuzzaman2024stase}, and regression analysis~\cite{IMPS}. These techniques can be leveraged to improve the efficiency of the summary generation component in our quantitative partial equivalence analysis as part of future work.


Although our range-search heuristic focuses only on numeric domains, numeric domains are relevant for a large set of vulnerabilities and corresponding patches. Moreover, our range-search heuristic can be extended to other domains (for example, by using alphanumeric ordering for strings). 

\section{Related Work}
\label{sec:relatedwork}

Prior research on patch assessment has primarily focused on identifying security patches and silent vulnerability fixes~\cite{sawadogo2022sspcatcher,tian2012identifying,zhou2021finding,wu2022enhancing,wang2019detecting,zhou2021spi}, as well as discovering vulnerability types~\cite{classifyPatches}. These approaches typically rely on machine learning models trained on commits, commit structures, code features, and extracted semantic representations to provide OSS users with more information about software updates. However, these methods do not account for the impact of updates across the input domain. Prior work on patching~\cite{spider} proposed a tool that automatically determines whether a patch is safe to deploy by analyzing only the original and patched code. However, a key limitation of their approach is that a patch may be deemed safe even if it restricts the entire input domain, a limitation explicitly acknowledged by the authors. In contrast, our approach can detect such cases by providing the equivalence and non-equivalence conditions between the original and patched code through a quantitative analysis of their input behavior.

Symbolic execution-based equivalence analysis has been explored in prior research~\cite{ardiff,IMPS,pasda}, focusing on formal proofs or refutations of equivalence between source codes. Differential symbolic execution~\cite{dse}, for instance, identifies functional differences using symbolic summaries, as our work does, but does not address partial equivalence with input conditions or quantification. Additional variations in equivalence analysis exist~\cite{ModDiff,clever,symdiff,shadowSym,shadowSym2}, including studies on compiler testing by examining divergent outputs across equivalent inputs~\cite{compilerValidation}. Equivalence under specific input conditions has also been studied~\cite{kawaguchi2010conditional}, which aligns with our notion of partial equivalence. However, none of these works focus on patch impact analysis with quantitative partial equivalence.



A related approach~\cite{quantificationSoftware1} investigates quantifying targeted software changes using incremental probabilistic symbolic execution~\cite{probabilisticSE}, employing model counting for quantitative analysis, though without addressing partial equivalence conditions on patch impact analysis. Previous work has also looked at quantifying the probability of reaching target events in programs via symbolic execution~\cite{iterativeProbabilisticSE,statististicalSESampling}. Additionally, model-counting-based quantification has been explored with tools like SMC~\cite{SMC}, S3\#~\cite{S3} for string domains, and LattE~\cite{latte} for linear integer arithmetic. Our work uniquely contributes a new range-search technique for partial equivalence analysis, which has not been explored previously.

Symbolic execution has been employed for vulnerability detection, patch correctness testing, and synthesis~\cite{shadowSEPatches,patchVulnerabilityDetechtionSE,patchTestingChoppedSE,correctnessCodeSE,patchSynthesisSE1,patchSynthesisSE2}. Additionally, studies on automated program repair leverage syntactic and semantic similarities between original and patched code~\cite{patchImpactSimilarity,patchImpactAnalysis,automatedPatchRepair,automatedPatchGeneration}. Our work, however, specifically targets patch impact analysis.

\section{Conclusion}
\label{sec:conclusion}

In this paper, we formalize the concept of partial equivalence for patch impact analysis and introduce a range-search–based approach for partial equivalence analysis. Our analysis shows that 36.25\% of program pairs in PatchBench—a benchmark derived from CVE patches—exhibit non-equivalent behavior over more than 90\% of the input domain. Additionally, our analysis distinguishes bad patches from good ones, as demonstrated using the Juliet Test Suite, with bad patches exhibiting greater non-equivalence relative to the original code. These findings underscore the importance of quantitative patch impact analysis.

Experimental evaluation on two benchmarks—EqBench (an equivalence analysis benchmark) and PatchBench—covering 118 program pairs shows that our proposed approach effectively computes equivalence conditions and provides sound lower bounds. It produces exact equivalence results in 86.2\% of cases with known ground truth, outperforming the next-best method by 30.7\%.

\appendix
\section{Appendix}
\begin{lstlisting}[caption=EqBench: ltfive,label={lst:eqBenchLtfive},language=C, xleftmargin=5.0ex,  escapeinside={(*@}{@*)},numbers=none, ]
int lib(int x){ (*@\aftergroup\speciallstcolorAdd@*)+ if(x < 5) return 5;(*@\aftergroup\endspeciallstcolor@*)
                (*@\aftergroup\speciallstcolor@*)- if(x < 0) return 0;(*@\aftergroup\endspeciallstcolor@*)
                  else return x;}
int client(int x){  if (x < 0) return -lib((-x)*5)/5;
                    else return lib((x+1)*5)/5-1; }
\end{lstlisting}

While analyzing EqBench benchmark using Algorithm~\ref{proc:classifier} for detecting partial equivalence, we found 4 totally non-equivalent ($T_{neq}$) programs. Among the partially equivalent cases detected, we found 5 cases where they are mislabeled as ``equivalent'' in the benchmark. Our SMT bit-vector–encoded summaries of program binaries enabled precise analysis to detect those cases. After  inspection, we determined that the reason behind the partial equivalence for those cases is integer overflow. For example, one of such cases is in Listing~\ref{lst:eqBenchLtfive}, where {\tt client} function calls {\tt lib} function with different values by comparing variable \texttt{x}. So, the two programs will be equivalent if they never reach the first branch in {\tt lib}. The multiplication of positive {\tt x+1} with $5$ in the second return can result in a negative value due to integer overflow and the same can happen in the first branch of {\tt client} function. Due to these integer overflows, the programs will generate two very different outputs.

\label{appendix}

\newpage
\bibliography{main}
\bibliographystyle{splncs04} 
\end{document}